\documentclass[aps,prapplied,reprint,
superscriptaddress,
 amsmath,amssymb,
]{revtex4-2}

\usepackage{graphicx}
\usepackage{dcolumn}
\usepackage{bm}
\usepackage{color}
\usepackage{wrapfig}
\usepackage[outline]{contour}
\definecolor{mblue}{RGB}{54,144,192}
\definecolor{mgreen}{RGB}{1,100,80}
\definecolor{mlgreen}{RGB}{2,120,138}
\definecolor{mlblue}{RGB}{103,169,207}
\definecolor{llblue}{RGB}{166,189,219}
\usepackage[colorlinks=true, linkcolor=mgreen, citecolor=mblue, urlcolor=mblue, breaklinks=true]{hyperref}

\usepackage[all]{hypcap}
\usepackage{siunitx}

\begin{document}


\title{\textbf{Liquid metal printing for superconducting circuits} 
}%

\author{Alexander Kreiner}
 \thanks{These authors contributed equally to this work.}

\author{Navid Hussain}
 \thanks{These authors contributed equally to this work.}
\affiliation{Institute of Nanotechnology, Karlsruhe Institute of Technology, Kaiserstraße 12, 76131 Karlsruhe, Germany}

\author{Ritika Dhundhwal}
\affiliation{Institute for Quantum Materials and Technologies, Karlsruhe Institute of Technology, Kaiserstraße 12, 76131 Karlsruhe, Germany}

\author{Haoran Duan}
\affiliation{Institute of Nanotechnology, Karlsruhe Institute of Technology, Kaiserstraße 12, 76131 Karlsruhe, Germany}

\author{Nicolas Zapata}
\affiliation{Institute for Quantum Materials and Technologies, Karlsruhe Institute of Technology, Kaiserstraße 12, 76131 Karlsruhe, Germany}

\author{Gabriel Cadilha Marques}
\affiliation{Institute of Nanotechnology, Karlsruhe Institute of Technology, Kaiserstraße 12, 76131 Karlsruhe, Germany}

\author{Tino Cubaynes}
\affiliation{Institute for Quantum Materials and Technologies, Karlsruhe Institute of Technology, Kaiserstraße 12, 76131 Karlsruhe, Germany}

\author{Torsten Scherer}
\affiliation{Institute of Nanotechnology, Karlsruhe Institute of Technology, Kaiserstraße 12, 76131 Karlsruhe, Germany}
\affiliation{Karlsruhe Nano Micro Facility (KNMFi), Karlsruhe Institute of Technology (KIT), Kaiserstraße 12, 76131 Karlsruhe, Germany}

\author{Wolfgang Wernsdorfer}
\affiliation{Institute for Quantum Materials and Technologies, Karlsruhe Institute of Technology, Kaiserstraße 12, 76131 Karlsruhe, Germany}
\affiliation{Physikalisches Institut, Karlsruhe Institute of Technology (KIT), Kaiserstraße 12, 76131 Karlsruhe, Germany}

\author{Michael Hirtz}
\affiliation{Institute of Nanotechnology, Karlsruhe Institute of Technology, Kaiserstraße 12, 76131 Karlsruhe, Germany}
\affiliation{Karlsruhe Nano Micro Facility (KNMFi), Karlsruhe Institute of Technology (KIT), Kaiserstraße 12, 76131 Karlsruhe, Germany}

\author{Ioan Mihai Pop}
\affiliation{Institute for Quantum Materials and Technologies, Karlsruhe Institute of Technology, Kaiserstraße 12, 76131 Karlsruhe, Germany}
\affiliation{Physikalisches Institut, Karlsruhe Institute of Technology (KIT), Kaiserstraße 12, 76131 Karlsruhe, Germany}
\affiliation{Physics Institute 1, Stuttgart University, Pfaffenwaldring 57, 70569 Stuttgart, Germany}

\author{Jasmin Aghassi-Hagmann}
 \email{Contact author: jasmin.aghassi@kit.edu}
\affiliation{Institute of Nanotechnology, Karlsruhe Institute of Technology, Kaiserstraße 12, 76131 Karlsruhe, Germany}

\author{Thomas Reisinger}
\affiliation{Institute for Quantum Materials and Technologies, Karlsruhe Institute of Technology, Kaiserstraße 12, 76131 Karlsruhe, Germany}


\date{\today}

\begin{abstract}
Superconducting circuits are a promising platform for implementing fault-tolerant quantum computers, quantum limited amplifiers, ultra-low power superconducting electronics, and  sensors with ultimate sensitivity. Typically, circuit fabrication is realized by standard nanolithography, generally associated with a high level of control over defects and contaminants. Additive approaches have not been used so far since they are expected to be inferior in terms of superconducting properties or quantum coherence. 
This work shows that liquid-metal based micro-pipette printing is suited for fabricating superconducting lumped-element resonators with high internal quality factors and thus its applicability for low-loss superconducting device fabrication. The possibility to locally add metal structures, without affecting any preexisting circuit elements, is a further advantage. Our results open up new avenues in the hardware implementation of scaled-up superconducting quantum computers.
\end{abstract}

\maketitle


\section{\label{sec:Intro}Introduction}

Quantum computing based on superconducting circuits has evolved rapidly throughout the past decade, culminating in recent demonstrations of quantum advantage in scaled-up quantum processors and quantum error-correction\cite{Krantz2019Jun, Blais2021cqed, Quantumsupremacy2019, Sivak2023}. However, decoherence remains a major obstacle to further improve 
quantum information processing fidelity, while achieving the required circuit sizes that would enable disruptive applications. Two complementary strategies are commonly pursued to enhance circuit coherence. The first aims to reduce susceptibility to the dominant loss channels, which can be achieved by engineering the circuit and its operating point so that coupling to these channels is minimized. The second seeks to suppress, eliminate, or passivate the underlying physical mechanisms that generate noise and dissipation.\cite{siddiqi_engineering_2021}. 
In superconducting qubits, such as transmons, coherence is often limited by dielectric loss arising at surfaces and interfaces of the superconducting capacitor electrodes\cite{Wang2015}. Fully encapsulating the superconductor, for example, with more noble superconducting metals, which would help to remove this loss channel by avoiding amorphous surface oxides, is a difficult task\cite{Mustafa2024FermiLab}. Increasing the size of the capacitor plates to reduce participation of the interfaces in the excited electro-magnetic mode is an alternative, which would fall into the first category mentioned above. However, conventionally, the circuits are realized using metallic thin films patterned using planar lithography on low-loss substrates. This technique results in unavoidable contamination from the fabrication steps required to produce and lift-off the patterning mask. Furthermore, the sharp edges of the thin film electrodes defined by the film thickness make the circuits particularly sensitive to dielectric loss incurred by amorphous surface oxides located there. A possible solution is therefore to fabricate the circuits using additive manufacturing approaches, such as printing, that are more suited to applying thick and round metal electrodes.

Techniques such as inkjet printing or drop casting are widely used for fabricating printed electronics\cite{Singh2010Feb, KaliyarajSelvaKumar2020Dec, Su2022Mar}. These methods have been also used to print superconducting circuits, for example, using ink dispersions based on superconducting NbSe2 flakes\cite{NbSe2_flakes} or 1T'-WS\textsubscript{2}\cite{1T-WS2_monolayer}. However, detailed reports of microwave loss for these materials are normally not available, particularly in the case for the low power excitation regime most relevant to quantum circuits\cite{McRae2020}. 
A promising class of materials for printed superconducting circuits is the use of low-melting point metals as ink, such as the eutectic alloy of gallium, indium, and tin (EGaInSn). These materials combine high electrical conductivity with fluidic deformability, making them especially attractive for soft and flexible electronics\cite{Rich2018Feb}. LMs have been patterned in diverse forms, for example, including bulk structures, in the form of nanodroplets\cite{EGaInSn_nanodroplets} or mixed with copper microparticles\cite{Cu-EGaInSn_brushing}. Recently, EGaInSn interconnects have also shown to not limit coherence in the circuits where the liquid metal was applied\cite{Yao2024Jun}.

At larger scales, additive processes such as selective laser melting of Nb powder or Al-12Si alloy have been used to fabricate low-loss superconducting 3D cavity resonators\cite{Nb_powder_cavity, Al-12Si_cavity}. Electron-beam powder-bed fusion has also been utilized to produce NbTi alloy 3D cavities\cite{NbTi_powder_cavity}. These cavities achieve quality factors of the order of millions when applying additional post‑processing steps, including surface polishing and heat treatments above \SI{200}{\degreeCelsius}. However, these are normally not easily adapted to planar circuits.

Here, we show that with an alternative printing technique, namely micro-pipette printing, superconducting lumped-element resonators with single-photon quality factors approaching 1 million can be fabricated using the liquid metal alloy EGaInSn. While this demonstrates the suitability of the technique for fabricating low-loss superconducting devices directly, we also find that the printed circuits degrade due to repeated thermal cycling, which we associate with a destructive phase transition occurring at low temperatures in addition to the expected solid/liquid phase transition. 
Engineering the constituents of the liquid metal, the substrate and shape of the resonator should in future allow for even higher quality factors and potentially help to avoid any phase transitions in the thermal cycling process.

\section{\label{Printing}Liquid metal printing}

Known for their liquid state near room temperature, liquid metals (LM) are processable through printing techniques while showing electrical and thermal conductivity similar to conventional metals\cite{Galinstan_resistivity}. In contrast to metal precursors, powders or nanoparticles, LM do not require additional processing steps involving annealing, sintering, or vacuum conditions. One major drawback of LM lies in a combination of viscosity, high surface tension, and density, posing a challenge to print them in the low-micrometer range with standard direct writing technologies\cite{Surface_tension, Surface_tension2}.

A solution to this problem was shown by Boley et al. using a syringe pump and a controllable stage to pattern eutectic gallium-indium (EGaIn) in micrometer scale\cite{Boley_LM_printing}. Since gallium oxidizes quickly when exposed to air, gallium-alloys form an outer oxide layer with a thickness of a few nanometer stabilizing the material when fabricated under environmental conditions\cite{Oxide_layer2, Oxide_layer1, Hussain2021Nov}. Probably the most common non-toxic liquid metal is Galinstan (brand name for EGaInSn alloy close to the eutectic composition), as it replaced mercury as a material for clinical thermometers.
Reports of the exact eutectic composition (e.g. 59.6~\% Ga, 26.0~\% In, 14.4~\% Sn (by weight)\cite{vanIngen1970Sep}) vary in the literature significantly\cite{Handschuh-Wang2021Sep,Handschuh-Wang2022Dec}, and we use the term EGaInSn to also refer to alloys similar in composition to the eutectic.
EGaInSn is already liquid at room temperature (melting point of about \qty{10}{\degreeCelsius}, depending on exact composition)
and displays a superconducting transition temperature around \SI{6}{K}, which is higher than the critical temperature of its individual components (\SI{1.08}{K} for Ga, \SI{3.41}{K} for In, and \SI{3.73}{K} for Sn)\cite{GaInSn_melting/freezing, EGaInSn_nanodroplets}. Interestingly, due to supercooling there can be a temperature offset of as much as \SI{30}{K} of the freezing point (\qty{-19}{\degreeCelsius}) from the melting point\cite{Handschuh-Wang2022Dec, GaInSn_melting/freezing, GaInSn_cooldown/warmup}.

\begin{figure*}
    \centering
    \includegraphics[width=\textwidth]{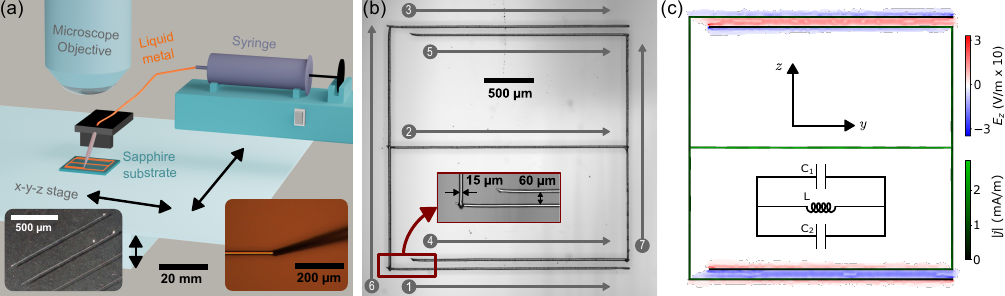}
    \caption{\textbf{Liquid metal capillary printing of superconducting lumped-element resonators.} \textbf{(a)} Illustration of the capillary printing setup used for fabricating superconducting resonators. The inset on the left shows a photograph of lines printed using this setup with the liquid metal alloy EGaInSn. The right-hand inset depicts the capillary tip during printing as seen through the optical microscope. 
    \textbf{(b)} Optical micrograph of an EGaInSn lumped-element resonator printed with the setup shown in (a). 
    The inset shows the lower left corner of the resonator at higher magnification, highlighting the achieved linewidth and pitch. The gray arrows specify the order and directions in which the resonator lines were printed.
    \textbf{(c)} Plots of the electric field magnitude $E_z$ (blue and red color scale) and magnitude of the current density $|j|$ (green color scale) derived from electro-magnetic finite-element eigen-mode simulation for the fundamental mode at $f_0 \approx \SI{5.5}{GHz}$. The plots correspond to the phases at which field and current are at a maximum, respectively. They show that the circuit can be interpreted as a lumped-element resonator with the equivalent circuit shown in the inset and $f_0 = 1/(2\pi\sqrt{L(C_1+C_2)})$. 
    } 
    \label{fig:printing}
\end{figure*}

\hyperref[fig:printing]{FIG.~1(a)} shows the microcapillary printing setup, which consists of a commercially available nanolithography platform that has been modified with a microfluidic syringe pump system\cite{Hussain2021Nov,Hussain2022Oct}. Commercially available EGaInSn (68\% Ga, 21\% In, and 11\% Sn by weight) was used as a liquid metal ink. A syringe acts as a reservoir for the liquid metal and is deployed into the holder of the pump system. The glass capillary, which is connected to the syringe via a rubber tube, is held in place by a stationary capillary holder and it is in contact with the substrate while printing. The system itself employs three piezo‐driven linear stages enabling movement in $x$-, $y$-, and $z$-direction and an optical microscope to facilitate the printing process. A photograph of four printed EGaInSn lines with varying linewidths can be found in the inset, highlighting the reflectivity of the material and the half cylinder-shaped structures that are result of this fabrication method.  

In this work, we apply the microcapillary LM printing technique to fabricate lumped-element resonators on sapphire substrates. As an example, an optical micrograph of a printed resonator (Resonator 1) is shown in \hyperref[fig:printing]{FIG.~1(b)}. The lateral size of the printed resonators is about \SI{2}{mm} $\times$ \SI{2}{mm}. The attained linewidths were just slightly above \SI{10}{\micro\meter} with a capacitance pitch of \SI{60}{\micro\meter}. 
The fabrication of LM resonators can be attained through a variety of different printing procedures. For the shown resonator, seven straight lines were printed in total in the order indicated by the gray numbered arrows. For each, the capillary tip disconnects from the substrate following the completion of each line. This approach offers simplicity, reproducibility, and a fabrication time of only a few minutes. An alternative printing approach was used to fabricate Resonators 2 and 3 (See Supplementary Section 1).

In \hyperref[fig:printing]{FIG.~1(c)} we  show plots from electro-magnetic simulations with the software Ansys HFSS, which we performed to predict resonator frequency and coupling. The simple lumped-element design can be described by  the shown equivalent circuit and is composed of two mostly capacitive elements at the top and bottom connected via the inductive trace in the middle. The plots indicate that for the fundamental mode the outer and inner electrodes are always oppositely charged and that the majority of the current within the resonator is found to flow through the inductive line element at the center, while only a negligible current is observed in the outer capacitive elements.

\section{\label{Characterization}Resonator characterization in a dilution cryostat}

\begin{figure*}
    \centering
    \includegraphics[width=\textwidth]{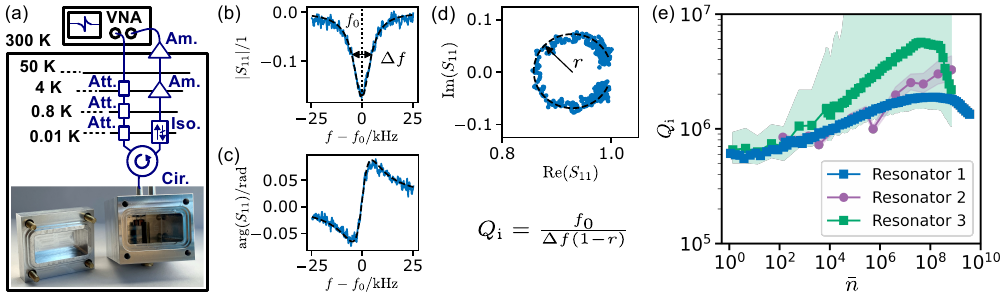}
    \caption{\textbf{Cryogenic measurement of the single-photon resonator quality factor $Q_\mathrm{i}$.} 
    \textbf{(a)} Schematic of the dilution cryostat measurement setup used to characterize microwave loss of the printed resonators at millikelvin temperature. 
    The resonators were mounted in an aluminum waveguide, as depicted in the photograph with the waveguide lid opened to reveal the resonator sample. It was attached to the mixing chamber stage of the cryostat.
    The plots in \textbf{(b)} and \textbf{(c)} show the reflected amplitude $|S_{11}|$ and phase $\mathrm{arg}(S_{11})$ measured with the vector network analyzer (VNA) for Resonator 1 near its resonance frequency $f_0 = \SI{5.791625}{GHz}$ and the lowest microwave power, resulting in resonator photon occupation $\bar{n}\approx1$ and thus most relevant for quantum applications. 
    In \textbf{(d)} the reflected signal is shown to follow a circular path in the complex plane. The dashed lines show the harmonic model fit for $S_{11}$ used to extract the internal quality factor $Q_\mathrm{i}$. The latter depends on the resonance width $\Delta f$ and circle radius $r$ as specified in the formula. 
    \textbf{(e)} Dependence of $Q_\mathrm{i}$ on the exciting microwave power, shown as the derived average photon number $\bar{n}$ stored in the resonator, for three printed resonators (Resonator 1 - 3). The shaded region indicates uncertainty from Fano interference\cite{Rieger2023Fano}.
    } 
    \label{fig:simulation_measurement}
\end{figure*}

The setup used to characterize the microwave properties of the printed resonators is shown in \hyperref[fig:simulation_measurement]{FIG.~2(a)}. The horizontal lines indicate the different cooling stages of the employed dilution cryostat with approximate temperatures noted in the labels. The sapphire chip with the printed resonator is placed in a microwave 3D waveguide made of aluminum as depicted in the photograph with the use of vacuum grease for thermalization.
The waveguide provides a clean microwave environment, which helps to avoid loss channels such as parasitic modes or vortex motion in the ground plane of co-planar waveguide setups, and therefore enables the measurement of the small levels of loss expected from highly coherent microwave circuit elements. 
The amplitude and phase of the reflected microwave signal are measured with a vector network analyzer (VNA). In order to achieve single-photon sensitivity, its microwave signal, which is swept in frequency, enters the cryostat via a chain of attenuators (Att.) to suppress noise from room-temperature electronics, and is directed at the impedance-matched waveguide using a circulator (Cir.). The signal reflected back from the waveguide exits the circulator via its third port, that is connected to an isolator (Iso.), suppressing noise back-radiated from the subsequent amplifiers (Am.), which the signal passes before returning to the VNA.
The resonator frequency $f_0$ and internal quality factor $Q_\mathrm{i}$ of the resonators are derived from microwave reflection measurements, which we perform by recording the complex $S_{11}$ coefficient with a vector network analyzer as a function of frequency.

The measured amplitude and phase of Resonator 1 at the lowest excitation power are depicted in \hyperref[fig:simulation_measurement]{FIG.~2(b, c)} together with a circle-based model fit\cite{qkit,Probst_analysis-of-complex-scattering-data}, as $S_{11}$ results in a circular path in the complex plane, as shown in \hyperref[fig:simulation_measurement]{FIG.~2(d)}. The off-resonant point in this plot is located at (1,0). Additional amplitude plots and circle fits for varying excitation powers can be found in Supplementary Section 2. Using the fit, we extract $f_0 = \SI{5.791625}{GHz}$, which is close to the frequency predicted by finite element simulation ($f_0 \approx \SI{5.5}{GHz}$), and $Q_\mathrm{i}$, where the model for the normalized and calibrated $S_{11}$ is given by the equation\cite{Probst_analysis-of-complex-scattering-data}:
\begin{equation}
    S_{11} = 1- \frac{2 Q_\mathrm{total} / Q_\mathrm{c}}{1+ 2iQ_\mathrm{total} (f / f_0 -1)} ,
\end{equation}
with the loaded quality factor $Q_\mathrm{total}$ and the coupling quality factor $Q_\mathrm{c}$. The internal quality factor $Q_\mathrm{i}$ characterizing the loss associated with the printed resonator itself is then derived from the equation: 
\begin{equation}
    \frac{1}{Q_\mathrm{i}} = \frac{1}{Q_\mathrm{total}} -\frac{1}{Q_\mathrm{c}}.
\end{equation}
Note that the total full width at half maximum (FWHM) $\Delta f$  of the amplitude is only about \SI{10}{kHz}, indicating a remarkably low level of loss. It is related to  $Q_\mathrm{total}$:
\begin{equation}
    \Delta f = \frac{f_0}{Q_\mathrm{total}}.
\end{equation}
In the under-coupled regime ($Q_\mathrm{c} > Q_\mathrm{i}$),  only a small phase roll is detected, and equivalently, a small radius of the circle in the complex plane. In this case the internal quality factor is nearly given by $Q_\mathrm{total}$, with only a small correction related to the circle radius $r=Q_\mathrm{total}/Q_\mathrm{c}$\cite{Rieger2023Fano}:
\begin{equation}
    Q_\mathrm{i} =  \frac{Q_\mathrm{total}}{(1-r)}.
\end{equation}
The under-coupled regime is therefore ideal for accurate measurements of microwave loss, which is the reason for choosing a resonator design here that results in a sufficiently large $Q_\mathrm{c}$ in the waveguide. Furthermore, the uncertainty of $Q_\mathrm{i}$ due to the Fano effect, which stems from the limited isolation of the circulator and the associated interference with the signal by-passing the waveguide, is less relevant in this regime\cite{Rieger2023Fano}.

In in \hyperref[fig:simulation_measurement]{FIG.~2(e)} the dependence of $Q_\mathrm{i}$ on incident microwave power $P_{in}$ is shown for printed Resonator 1, 2, and 3 during the first cool-down at a base temperature of approximately \SI{10}{\milli\kelvin}. In the graph, $Q_\mathrm{i}$ is plotted against the average photon number $\bar{n}$, which is determined from $P_{in}$ using the equation

\begin{equation}
    \bar{n} = 4 \frac{Q_\mathrm{total}^2}{Q_c \hbar (2\pi f_0)^2} P_{in}.
\end{equation} 
$P_{in}$ is estimated by subtracting the attenuation on the input line from the nominal output power of the VNA at room temperature. The shaded areas in the plot indicate Fano uncertainty of the $Q_\mathrm{i}$ measurement. In the single photon regime, which is most relevant to quantum applications, we consistently find $Q_\mathrm{i}\approx 6\times 10^5$, comparable to standard implementations using lithographically patterned aluminum or tantalum.

\begin{figure*}
    \centering
    \includegraphics[width=\textwidth]{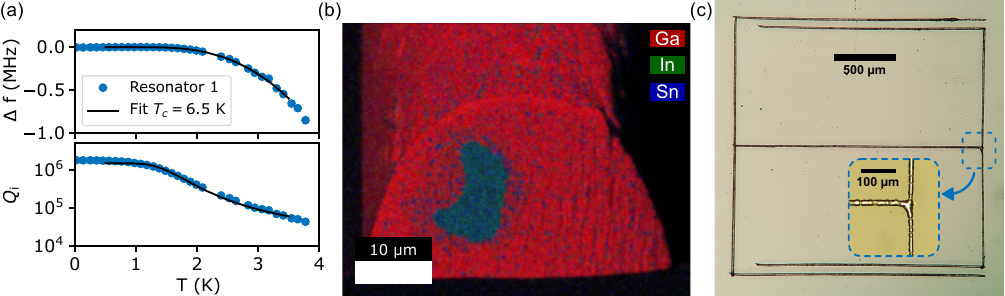}
    \caption{\textbf{Superconducting critical temperature and effects of cool-down and warm-up on Resonator 1.}
    \textbf{(a)} Frequency shift $\Delta f(T) = f(T)-f_0$ and quality factor $Q_\mathrm{i}(T)$ for Resonator 1 as a function of temperature. The solid lines are a fit to a model for surface impedance due to thermal quasi-particles yielding EGaInSn's superconducting critical temperature $T_\mathrm{c}$.
    \textbf{(b)} Energy-dispersive X-ray elemental map of the cross-section of a printed EGaInSn line at around \SI{88}{\kelvin} reveals a segregation of gallium (red) from indium (green) and tin (blue). 
    \textbf{(c)} Optical micrograph of Resonator 1 after the first cryogenic measurement. Markedly, the uppermost printed trace has changed in length likely due to dewetting. Where the liquid metal has retracted, some residues remain which have possibly alloyed with the substrate. The inset shows that the printed traces have increased in surface roughness with intermittent constrictions. 
    } 
    \label{fig:resonatorsurvives}
\end{figure*}

In addition to characterizing $Q_\mathrm{i}(\bar{n})$, in \hyperref[fig:resonatorsurvives]{FIG.~3(a)} the change in frequency $\Delta f(T) = f(T)-f_0$ and $Q_\mathrm{i}(T)$ with temperature is shown. In the plot, the solid black lines show the result of fitting both quantities jointly to an analytical model for thermal quasi-particle excitations of the superconductor\cite{Crowley2023disentangling_losses_Ta}. In addition to  $T_\mathrm{c}$, the following parameters were determined from the fit: kinetic inductance fraction $\alpha = 0.14~\%$, a scaling factor for thermal quasi-particle loss $A_{QP}=324$, and $Q_{\mathrm{other}}=1.5\times 10^6$, which likely sets the level of dielectric losses. The determined value of $T_\mathrm{c}=\SI{6.5}{K}$, coincides with previously reported critical temperatures of EGaInSn\cite{GaInSn_Tc6K, EGaInSn_nanodroplets, GaInSn_cooldown/warmup}.

In order to determine how this effective superconducting critical temperature can be attributed to the various components of the inhomogeneous microstructure of solid EGaInSn, we performed Energy Dispersive X-ray Spectroscopy (EDS) mapping of the cross-section of a printed trace. The result is shown for a temperature of about \SI{88}{\kelvin} (the lowest temperature available in the setup) in \hyperref[fig:resonatorsurvives]{FIG.~3(b)}. The image shows the distributions for each element encoded as its red (gallium), green (indium) and blue (tin) component. The brightness of each component indicates the relative concentration within the alloy. The map reveals a distinct segregation of gallium (with finely dispersed, micron-sized patches of indium and tin) from a large indium- and tin-rich region inside the trace. The extracted relative concentrations of the latter region allow us to associate it with In\textsubscript{3}Sn. The literature values in the range $T_\mathrm{c}=\SIrange{6.6}{7}{\kelvin}$ for the superconducting critical temperature of this phase corresponds well to the one derived here from the thermal quasi-particle model \cite{Wernick_In3Sn_Tc, Bartram_In3Sn_Tc}.
For the $T_\mathrm{c}$ of the gallium-rich region we expect a similar value under the assumption that gallium is present in its $\beta$-phase ($T_\mathrm{c}~=~\SIrange{5.9}{6.2}{K}$)\cite{Moura_BetaGallium_Tc, BetaGallium_Tc, Quan_BetaGallium_Tc}. Note that the microstructure at low temperature likely depends on the cooling rate to some extend\cite{Bartram_In3Sn_Tc}, and may thus differ when cooling the samples in the dilution cryostat. 
Similar EDS measurements at higher temperatures reveal that the phase separation occurs already close to the solidification temperature and disappears after re-melting the trace (See Supplementary Section 3).

In \hyperref[fig:resonatorsurvives]{FIG.~3(c)} Resonator 1 is shown re-melted after the full cryo cycle. Apart from a small part of the outer capacitor line at the top right corner, the resonator is structurally intact. Upon closer inspection, changes in the LM reflectivity and small notches in the lines are visible. The exact origin of this dewetting process is still unknown but it is assumed that it occurs at the phase transition from solid to liquid around the melting point of EGaInSn (\qty{10}{\degreeCelsius}). Nonetheless, the structure is still intact and can be measured again during a second cool-down. 
Unfortunately, $Q_\mathrm{i}$ of the resonator could not be measured at this frequency as the resonator hybridized with a box-mode of the waveguide (see Supplementary Section 4). This box mode then dominated the loss rate, resulting in a combined $Q_\mathrm{i}\approx 2\times 10^4$, representing a lower bound for the printed resonator's $Q_\mathrm{i}$.

Several other resonators have been fabricated on intrinsic silicon and magnesium oxide (MgO). Supplementary Section 5 provides optical micrographs and $Q_\mathrm{i}$ measurements of these resonators. While the results demonstrate that the printing method is compatible with other substrates, only for silicon similarly high quality factors could be demonstrated as for sapphire.

\section{\label{Microscopy}Optical cryo-microscopy}

\begin{figure*}[t]
    \centering
    \includegraphics[width=\textwidth]{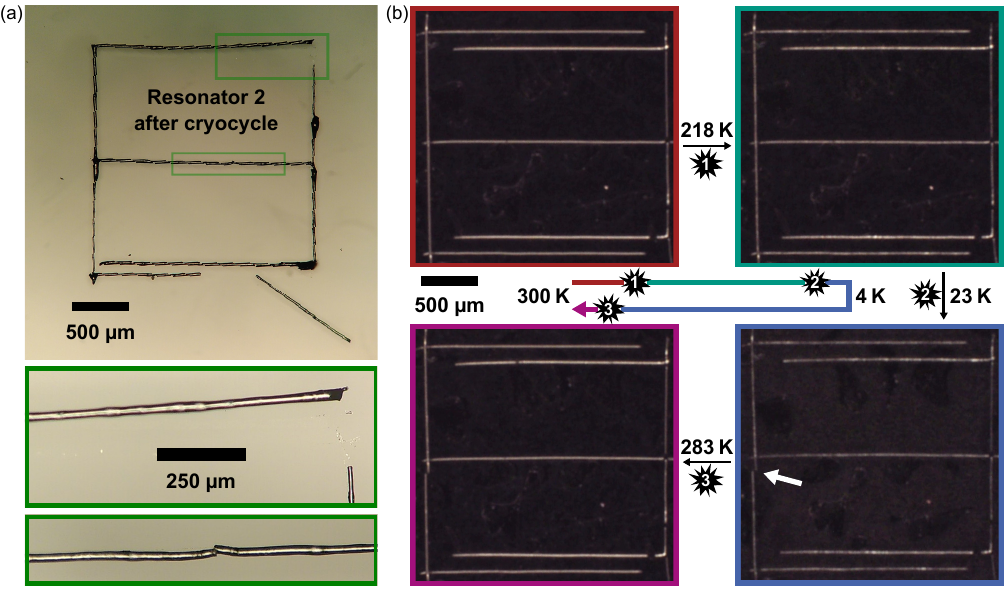}
    \caption{ \textbf{Destructive effect of thermal cycling on printed traces.} 
    \textbf{(a)} Optical micrograph of Resonator 2 after first cool-down/warm-up. 
    The insets show affected parts of the resonator at higher magnification. Some lines were fully detached from the substrate while at other locations more local defects occurred.
    \textbf{(b)} Optical cryo-microscopy experiment with a printed resonator. This was conducted in a separate helium flow cryostat probe station with an optical access window, in order to observe the printed structure during cool-down and warm-up. The left-most image shows the full resonator structure at room-temperature as mounted in the cryostat. 
    The colored segments indicate the temperature ranges in which the recorded micrographs did not change. Changes associated with the solid/liquid phase transformation occurred at \SI{218}{K} and \SI{283}{K} and a destructive change was observed during the cooling at about \SI{23}{K}.  
    } 
    \label{fig:breakingresonators}
\end{figure*}

In contrast to Resonator 1, the shape of Resonators 2 and 3 was affected by the cryo cycle more severely, with some parts of the resonator traces breaking, as can be seen in \hyperref[fig:breakingresonators]{FIG.~4(a)}. The sharp break points and disappearance as well as displacement of straight traces indicate that the temperature at which the resonators break is below the melting temperature of the alloy.

In \hyperref[fig:breakingresonators]{FIG.~4(b)} we show the result of investigating the cause for the structural failure by optical cryo-microscopy. A printed resonator sample was mounted in a cryogenic probe station equipped with an optical access port and microscope for imaging. The sapphire chip was attached to the cryogenic stage using vacuum grease and placed on top of a silicon chip to improve imaging contrast. While cooling the sample to about \SI{4}{\kelvin} at a maximum rate of about \SI{10}{K.min^{-1}}
and allowing it to warm up back to room temperature at a rate of about \SI{1.5}{K.min^{-1}}, 
a video was recorded. The cooling and warming rates are visualized in Supplementary Section 6. We could observe changes in the resonator at the indicated cryostat temperatures.  The images show the microscopy images before and after these transitions. Videos showing the full evolution of the microscopy images at the transitions are provided in the supplementary information (Supplementary Videos 1–3).

The first visible change in the EGaInSn resonator occurs around \SI{218}{K}, where the printed traces become less homogeneous in linewidth and reflectivity, indicating a roughened surface. We expect the actual temperature of the metal to be higher by at least a few kelvin due to the fast cooling rate and limited thermalization. As the used alloy deviates from the eutectic composition, we expect there to be no single solidification temperature. The observed transition most likely corresponds to one of the solidification temperatures of EGaInSn which is known from literature to vary significantly from the melting point, reaching values as low as \SIrange{226}{263}{\kelvin} associated with supercooling or solid-solid phase transitions\cite{Handschuh-Wang2021Sep,Handschuh-Wang2022Dec,Cu-EGaInSn_brushing}. Furthermore, close to or during the solidification the alloy decomposes, as discussed in the EDS mapping\cite{GaInSn_Tc6K}. In the video the whole resonator transforms around \SI{218}{K}  within \SI{200}{\milli\second}, and during this time the various smaller sections transform from one video frame to the other, i.e. less than \SI{6}{\milli\second} (See Supplementary Video 1). 

The melting transition of the alloy is also visible during the warm-up, as the traces become smoother and more reflective around \SI{283}{K}, which is close to the expected melting temperature of EGaInSn. The change at this transition occurs more gradually compared to the solidification during a time of about one minute (See Supplementary Video 3).

A third transition occurs around \SI{23}{K}, which causes the resonator to break. At this temperature, a pronounced change in reflectivity of the LM can be observed taking about two seconds in total, which starts at various locations and the phase boundary can be seen to propagate along the LM traces at a speed of about \SI{4}{\milli\meter\per\second}. Towards the end, the resonator clearly moves and loses contact with the substrate in some places. One of the traces even breaks at the location indicated by the white arrow. Considering the large change in shape, associated with a volume expansion most clearly seen in the trace second from the bottom of image (See Supplementary Video 2), it is likely that this phase transition is related to the tin pest, which is observed in many tin-based alloys\cite{Cornelius_tin_pest}. The likelihood of this allotropic phase transition in tin rises with decreasing temperature and can be promoted by low-solubility inclusions of for example aluminum\cite{Cornelius_tin_pest}. We note that most resonators printed on sapphire substrates, except for Resonator 1, broke during thermal cycling, while we did not observe the same pronounced breaking behavior on other substrates. Diffusion of aluminum from the sapphire is a probable explanation for this. Since Resonator 2 and 3 could be excited at close to the expected microwave frequency at temperatures below \SI{1}{\kelvin}, we deduce that the destructive transition did not occur for these samples during cooling but rather during warm-up.

\section{\label{Con}Conclusion}

We used liquid-metal based micro-capillary printing to fabricate superconducting lumped-element resonators with linewidths close to \SI{10}{\micro\meter} and single-photon quality factors approaching 1 million. This demonstrates the suitability of the technique for fabricating low-loss superconducting devices without any post-processing needed. One particularly promising application is to locally add superconducting circuit elements, without affecting any structures formed nearby in preceding steps. 
The superconducting critical temperature determined from frequency shift and thermal quasi-particle loss as a function of temperature is about $T_\mathrm{c} = \SI{6}{K}$, corresponding well to previous reports in the literature for the used GaInSn alloy.  The printed resonators are found to degrade due to thermal cycling, often in a catastrophic fashion. Optical cryo-microscopy and EDS gives strong evidence that this effect is associated with a destructive phase transition occurring at low temperature, likely corresponding to the well-known tin pest that many tin-based alloys suffer from, which we expect can be avoided in the future by choosing suitable additives to EGaInSn or alternative low-melting point alloys.

\begin{acknowledgments}
The authors are grateful to L. Radtke, B. Breitung and A. Khandelwal 
for technical support. This work was partly carried out with the support of the Karlsruhe Nano Micro Facility (KNMFi, www.knmf.kit.edu), a Helmholtz Research Infrastructure at the Karlsruhe Institute of Technology (KIT, www.kit.edu). Facilities use was supported by the Karlsruhe Institute of Technology (KIT) Nanostructure Service Laboratory (NSL). We acknowledge qKit for providing a convenient measurement software framework.

This project has received funding from the Helmholtz Association, the European Union’s Horizon 2020 research and innovation programme under the Marie Skłodowska-Curie grant agreement number 847471, the Federal Ministry of Education and Research (Projects QSolid (FKZ:13N16151, FKZ 13N16159) and GeQCoS (FKZ: 13N15683)). In addition, J.A., H.D. and A.K. acknowledge funding by the Joint-Lab Virtual Materials Design initiative and the Deutsche Forschungsgemeinschaft (DFG, German Research Foundation) under Germany's Excellence Strategy via the Excellence Cluster “3D Matter Made to Order” (EXC-2082/1-390761711), which has also been supported by the Carl Zeiss Foundation through the “Carl-Zeiss-Foundation Focus$@$HEiKA”, by the State of Baden-Württemberg and KIT.
\end{acknowledgments}

\section*{Data availability}
The data that support the findings of this study are available in the KIT Open Repository at:

http://doi.org/10.35097/zj91nubn3q790p8h.

\appendix

\section{Sample preparation and printing procedure}

\subsection{Preparation of Liquid Metal Ink} EGaInSn (\qty{56.4}{\percent} gallium, \qty{25.6}{\percent} indium, and \qty{18}{\percent} tin alloy by weight) was purchased from Strategic Elements, Germany, and used as received.

\subsection{Preparation of Substrates} Sapphire substrates were subjected to a sonic cleaning procedure involving acetone, 2-propanol, and deionized water, and dried by blowing with nitrogen.

\subsection{Preparation of Nozzles} A Sutter Instruments P-1000 micropipette puller system was used to prepare glass capillaries. Glass pipette without filament (outside diameter: \SI{1.2}{mm}, inside diameter: \SI{0.94}{mm} and overall length: \SI{10}{mm}) from Warner Instruments (GC120TF-10) were used. Pulling parameters for heat (600), pull (90), velocity (vel: 100), time (250), no delay, and pressure (500) were used, which provided a tip around \SI{3}{\micro m}, and taper 6-\SI{8}{mm}. The tip openings were manually enlarged to around \SI{20}{\micro m} by breaking the taper. 

\subsection{Liquid Metal Printing} LM printing was performed on an NLP 2000 system (Nanoink, USA) with a self-made holder for attaching the printing capillary\cite{Hussain2021Nov}. Generally, substrate treatments producing a more hydrophilic surface, like plasma cleaning or sonication in DI Water, were found to be favorable for printing EGaInSn. The substrates are fixed to the NLP 2000 sample stage by either using magnets or applying a droplet of water between the substrate and the printing stage. In order to print consistent liquid metal lines onto a surface, it is necessary to establish a constant liquid metal flow from the syringe to the glass capillary, balancing the speed of lateral movement (printing speed) of the capillary in relation to the substrate. This conditioning is done on a sacrificial area of a substrate by manually adjusting the pressure exerted on the LM reservoir by the flow rate settings of the syringe pump. Applying too much pressure will result in the formation of liquid metal droplets, while too little pressure results in discontinued lines. Usually, applying a flow rate of \SI{90}{\micro l.min^{-1}} was found to be sufficient for forming continuous lines with the fabricated capillary tips. Once a continuous flow of LM printing is achieved, the syringe pump is stopped, and the pressure remaining in the LM reservoir is sufficient to print multiple resonator structures, each taking around \SIrange{5}{10}{mins}. When LM flow is decreasing, the pressure in the LM reservoir is replenished by briefly activating the flow rate of the syringe pump until the flow stabilizes and the next couple of resonators can be printed. 

Important printing parameters are linewidth, line length, and the printing speed. The width of a line is largely dependent on the opening size of the capillary tip. The user interface of the NLP 2000 allows for the adjustment of line length and the printing speed. Varying the printing speed between \SIrange{200}{1000}{\micro m.s^{-1}} also shows a slight impact to the linewidth, where speeds at the higher end of the spectrum lead to moderately thinner lines. Moreover, printing at higher rates increases writing reliability and reproducibility since the formation of EGaInSn spheres is more likely at lower rates\cite{Hussain2022Oct}.

\section{Alternative printing procedure}
\begin{figure*}
    \centering    
    \includegraphics[width=\textwidth]{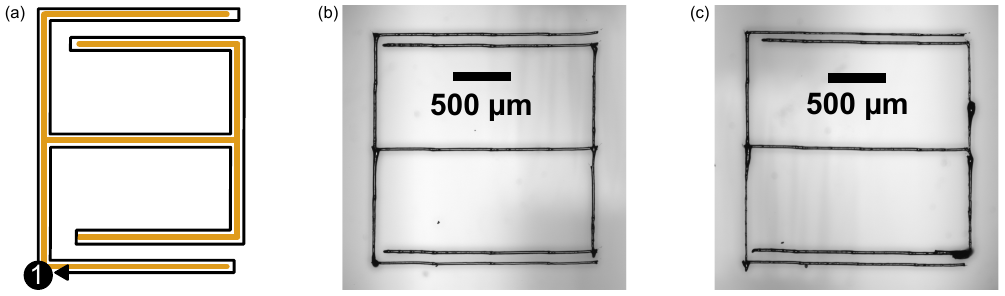}
    \caption[Printing path followed to fabricate Resonators 2 and 3.]{\textbf{Printing path followed to fabricate Resonators 2 and 3.}
    The traces of Resonators 2 and 3 were printed following a path as depicted in \textbf{(a)}, which is different to the multi-path approach followed for Resonator 1.
    The optical micrographs in \textbf{(b)} and \textbf{(c)} show Resonators 2 and 3, respectively.
    } 
    \label{fig:EDF_1_Printing}
\end{figure*}

The Resonators 2 and 3 were fabricated using an alternative printing procedure, which is highlighted in \hyperref[fig:EDF_1_Printing]{FIG.~5}. In this approach, the substrate and capillary tip remain in contact during the entire printing process. The designed resonator structure cannot be fully printed in a single pass without lifting the capillary. Therefore, in order to ensure galvanic contact we decided to retrace the lines, eventually returning to the initial point. The selected printing path is as shown with common start and end point. In this approach, the lines were printed in \SI{100}{\micro m}-steps. At the corners of the resonator, the tip was lifted by approximately \SI{5}{\micro m} from the substrate and then lowered again by the same amount to avoid any drifting of the capillary tip. This is done similarly when retracing the previously printed lines. The capillary holder allows to adapt the contact angle in the printing direction (x-direction), which is usually set to roughly \SIrange{50}{60}{\degree}, permitting good adhesion of the liquid metal to the substrate. Furthermore, printing in y-direction, which was done in this approach, is more difficult. It negatively impacts line reproducibility and makes discontinued lines more likely, since the contact angle in y-direction is fixed to \ang{90}, and cannot be adjusted. When retracing, the tip was lifted by around \SI{10}{\micro m} (not disconnecting capillary flow) to prevent displacing the already printed line. With this mostly manual approach the yield of the printed resonators was low. However, for Resonators 2 and 3 the result was reasonable, and the microwave measurements indicate that printing the lines separately also yields galvanically connected traces.

\section{Microwave measurements}
\begin{figure*}
    \centering
    \includegraphics[width=\textwidth]{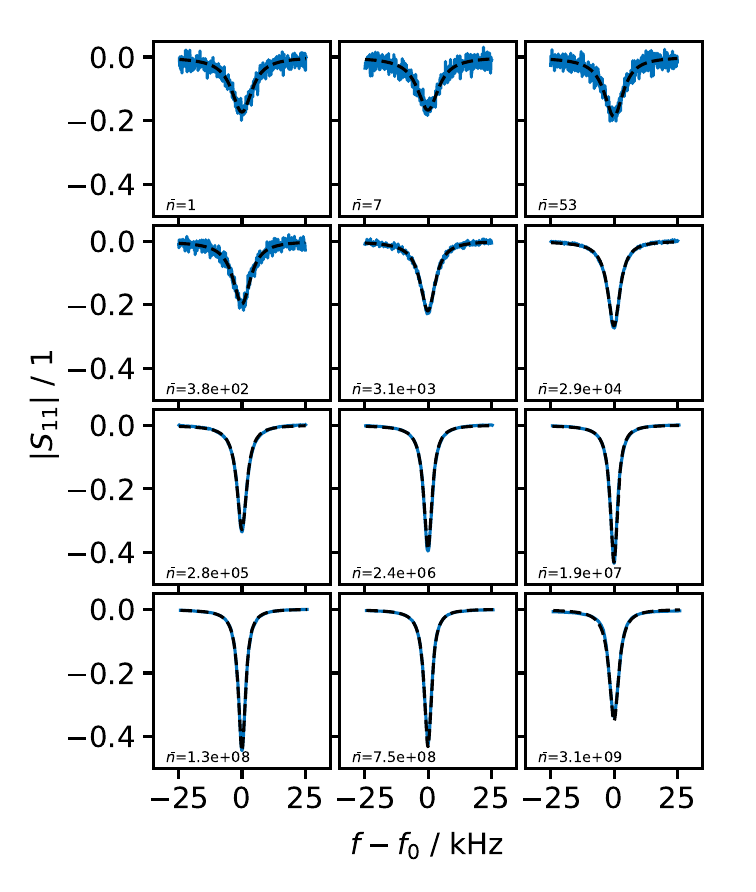}
    \caption[Power dependence of the amplitude line-shape.]{\textbf{Power dependence of the amplitude line-shape.} The reflection amplitude profiles of the fundamental mode of Resonator 1 are shown at different input powers, which are given as average photon numbers. The width of the line-shape indicates the change in $Q_\mathrm{i}$ with power, as reported in the main text.  The dashed lines show the harmonic model fit for $S_{11}$.
    } 
    \label{fig:EDF_2_Amp_Sweep}
\end{figure*}

\begin{figure*}
    \centering
    \includegraphics[width=\textwidth]{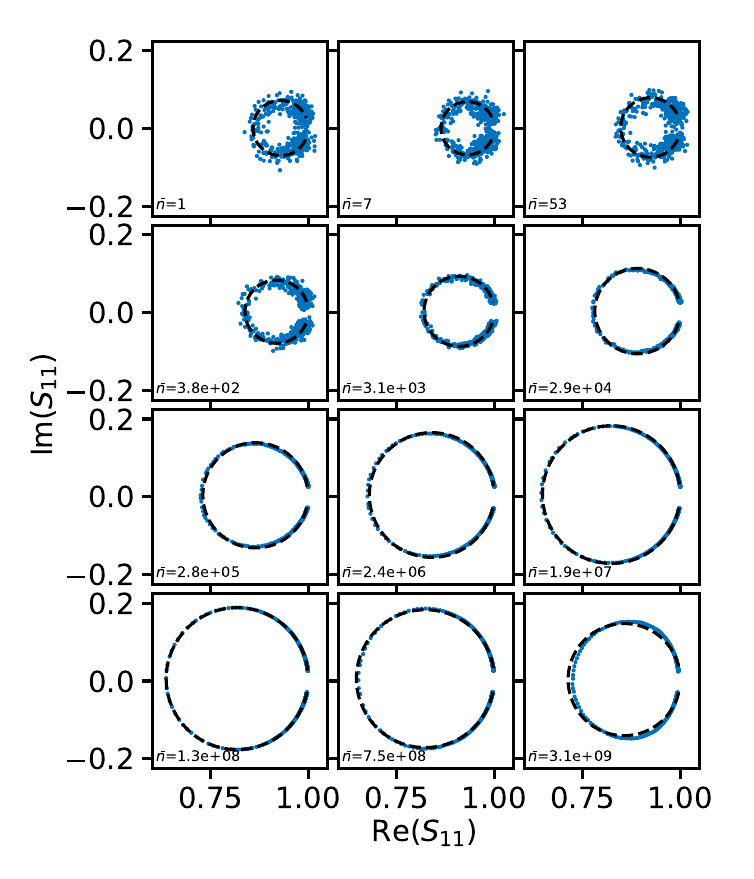}
    \caption[Power dependence of complex reflection coefficient for Resonator 1.]{\textbf{Power dependence of complex reflection coefficient for Resonator 1.} In addition to the experimental data shown as blue dots, the dashed lines show the harmonic model fit for $S_{11}$ corresponding to circles of varying diameter. The input powers are given as average photon numbers. Note that these only apply on resonance, and the occupation decreases for more off-resonant points. Therefore, the power dependence of loss leads to small deviations from the perfect circular shape. First it elongates the circle in the horizontal axis, as loss decreases with power, then this trend reverses leading to a horizontal compression of the circle.  }
    \label{fig:EDF_3_Circle_Sweep}
\end{figure*}

\hyperref[fig:EDF_2_Amp_Sweep]{FIG.~6} shows the reflected amplitude of the fundamental mode of Resonator 1 for increasing microwave input power, each given as average photon numbers $\bar{n}$. In \hyperref[fig:EDF_3_Circle_Sweep]{FIG.~7} the reflection coefficient is shown in the complex plane for the same microwave input powers together with the circle fit that was used to extract external and internal quality factors. Up to $\bar{n} = 1.3e+08$ the internal quality factor increases with power, resulting in a narrower line shape and a larger circle radius. This is due to a saturable loss mechanism, possibly associated with a the saturation of a two-level system bath. Increasing the microwave input power further, the trend is reversed, likely because a different loss mechanism that increases with power, such as quasiparticle heating, starts to dominate.  


\begin{figure*}
    \centering
    \includegraphics[width=\textwidth]{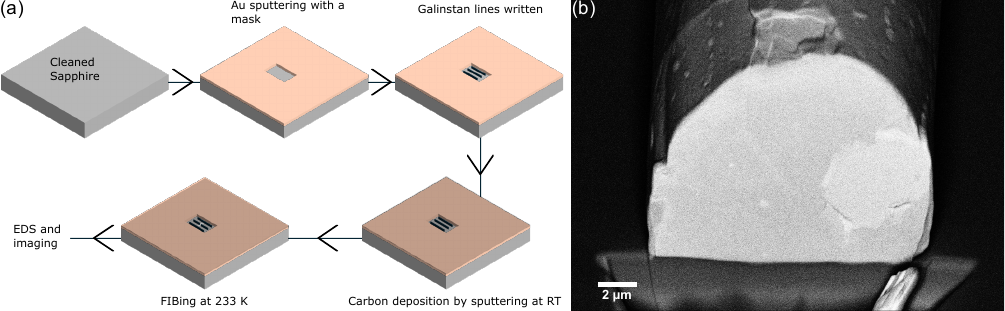}
    \caption[Energy Dispersive X-ray Spectroscopy (EDS) sample preparation and Backscattered Electron (BSE) image.]{\textbf{Energy Dispersive X-ray Spectroscopy (EDS) sample preparation and Backscattered Electron (BSE) image.} 
    \textbf{(a)} Schematic showing the preparation steps of the sapphire substrate that was used to reduce charging effects during the EDS measurement.
    \textbf{(b)} BSE image recorded at \SI{233}{\kelvin} showing the cross-section of the EGaInSn line printed on a sapphire substrate.
    } 
    \label{fig:EDF_4_Sample}
\end{figure*}

\section{EDS measurements}
In order to investigate the phase separation known to occur in EGaInSn, we performed Energy Dispersive X-ray Spectroscopy (EDS) mapping of the cross-section of a printed trace prepared by focused-ion beam milling. The sapphire sample preparation is shown in \hyperref[fig:EDF_4_Sample]{FIG.~8}. As sapphire is insulating, Gold and Carbon sputtering are necessary since charging effects can be a serious problem. The included Backscattered Electron (BSE) imaging revealed the onset of phase separation within the liquid metal line cross-section.

\begin{figure*}
    \centering
    \includegraphics[width=0.8\textwidth]{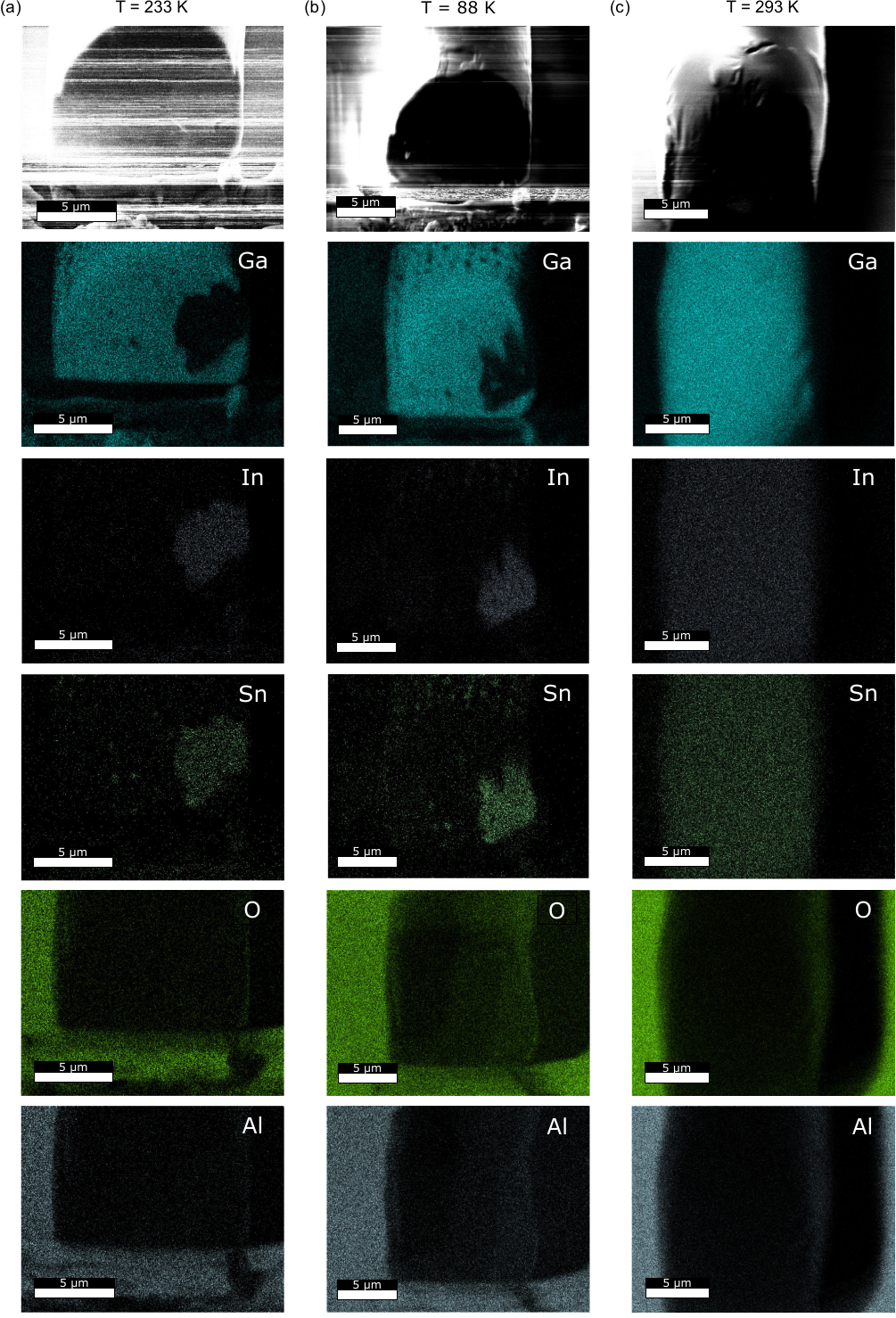}
    \caption[Energy Dispersive X-ray Spectroscopy (EDS) mapping of a printed EGaInSn line on sapphire.]{\textbf{Energy Dispersive X-ray Spectroscopy (EDS) mapping of a printed EGaInSn line on sapphire.} Scanning Electron Microscope (SEM) image of the Focused Ion Beam (FIB) cut EGaInSn line 
    and distribution of the individual elements gallium, indium, tin, oxygen, and aluminum within the cross-section at \textbf{(a)} \SI{233}{K}, \textbf{(b)} \SI{88}{K}, and \textbf{(c)} \SI{293}{K}.
    } 
    \label{fig:EDF_5_Sapphire}
\end{figure*}

The EDS mapping, as presented in \hyperref[fig:EDF_5_Sapphire]{FIG.~9}, was conducted at \SI{233}{K}, \SI{88}{K} and \SI{293}{K} (after temperature cycling). As can be seen in the three SEM images, charging effects could not be suppressed entirely. Therefore, the imaging had to be interrupted when too high drifting occurred  resulting in drift correction failure. This limited the signal-to-noise ratio in the images.
The corresponding elemental distributions are shown for gallium, indium, tin, oxygen and aluminum, respectively. In particular, gallium (Ga) is segregated distinctly from a large indium (In) and tin (Sn) rich region located fully inside the trace in addition to approximately micron sized patches finely dispersed throughout the Ga-rich phase. This early evidence of compositional decoupling is consistent with the morphological changes previously observed in the optical microscope at \SI{218}{K}, suggesting that the redistribution of metallic phases begins to manifest at relatively moderate cryogenic temperatures. The same area was remapped after cooling to \SI{88}{K}. At this lower temperature, EDS mapping revealed only minor changes in the phase distribution, remaining similar to that at \SI{233}{K}. 

To investigate the reversibility of this behavior, the sample was gradually warmed back to room temperature, with the temperature controller set to \SI{310}{K}, and mapping performed after thermal stabilization, beginning at \SI{293}{K}. The follow-up EDS mapping clearly illustrates that the previously observed segregation of the liquid metal components had disappeared, and the elemental distribution returned to a homogeneously intermixed state. 

\begin{figure*}
    \centering
    \includegraphics[width=0.8\textwidth]{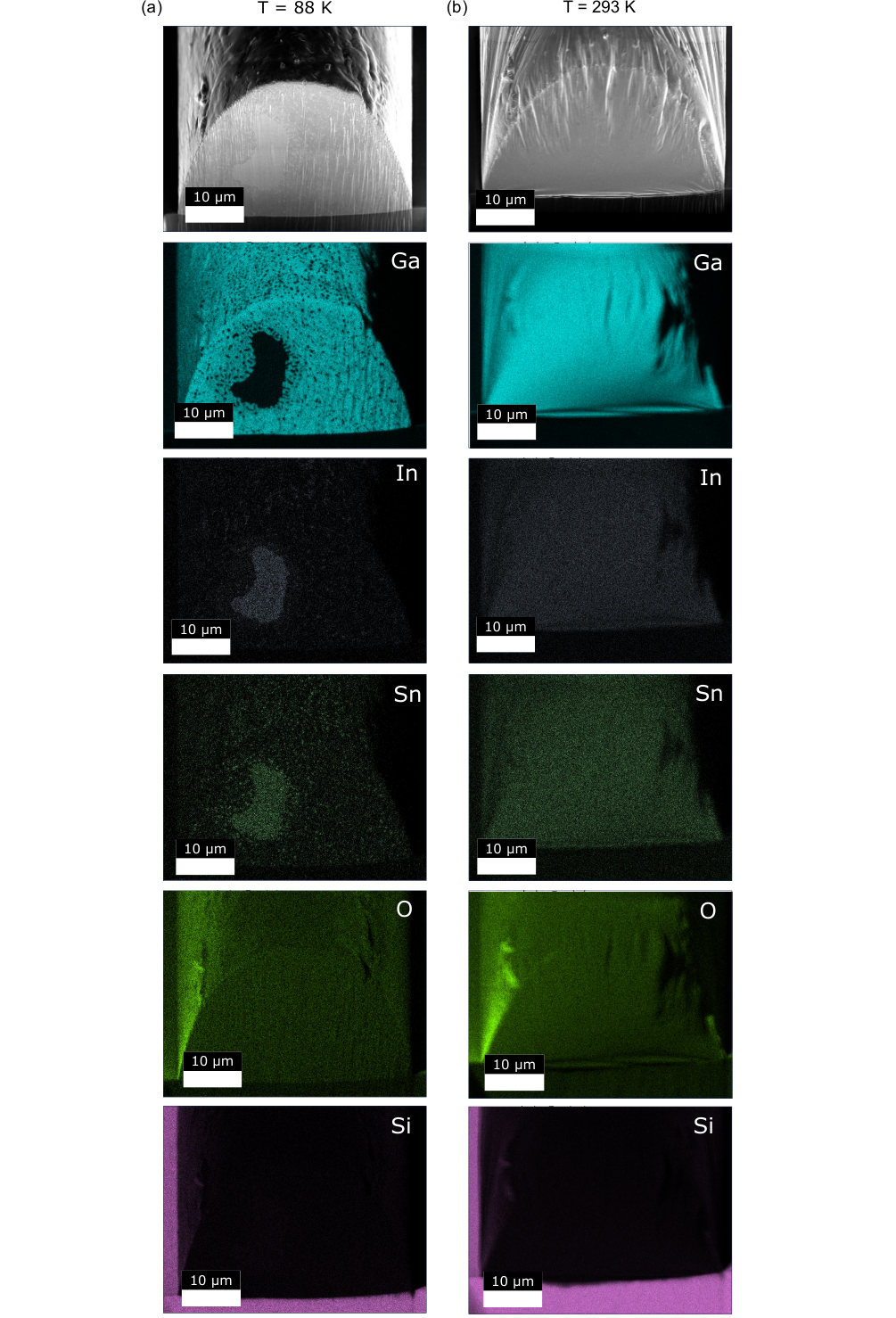}
    \caption[Energy Dispersive X-ray Spectroscopy (EDS) mapping of a printed EGaInSn line on silicon.]{\textbf{Energy Dispersive X-ray Spectroscopy (EDS) mapping of a printed EGaInSn line on Silicon.} Scanning Electron Microscope (SEM) image of the Focused Ion Beam (FIB) cut EGaInSn line 
    and distribution of the individual elements gallium, indium, tin, oxygen, and silicon within the cross-section at \textbf{(a)} \SI{88}{K}, and \textbf{(b)} \SI{293}{K}.
    } 
    \label{fig:EDF_6_Silicon}
\end{figure*}

Due to the high drifting encountered on the sapphire substrate, another sample was prepared on a silicon substrate. The EDS results can be seen in \hyperref[fig:EDF_6_Silicon]{FIG.~10}.   Here, the SEM images show no charging artifacts, and readily reveal a similar phase separation within the line cross-section and at higher contrast. The Ga, In and Sn maps were used to generate the elemental map in Figure~3(b) of the main text.
The oxygen map shows that there are no inclusions of oxide and the silicon map does not show any inter-diffusion between EGaInSn and the substrate. The phase separation disappears after re-melting the trace.

\subsection{Scanning Electron Microscopy Imaging, FIB and Energy-Dispersive X-Ray Spectroscopy} All imaging and compositional analyses were carried out using a Zeiss AURIGA 60 CrossBeam Focused Ion Beam\slash Scanning Electron Microscope (FIB/SEM) system equipped with a GATAN temperature control stage, which enabled cooling down to cryogenic temperatures as low as \SI{88}{K} using liquid nitrogen. The FIB column, equipped with gallium ion Ga$^{+}$ sources, was employed for site-specific material removal and preparation of cross-sections. Milling was performed using an acceleration voltage of \SI{30}{kV} with a beam current of \SI{4}{nA} to achieve rapid material removal, followed by a final polishing step at \SI{30}{kV} and \SI{1}{nA} to smooth the milled cross-section surfaces and minimize surface damage or redeposition artifacts.

SEM imaging was conducted using the field emission electron column of the same instrument. High-resolution images were acquired under a primary electron beam energy of \SI{5}{keV} and inLens detector.

Elemental analysis and mapping were performed using the same Zeiss AURIGA 60 system, integrated with an EDAX Octane Super detector, operated via the EDAX TEAM software suite. For EDS measurements, an electron beam with an accelerating voltage of \SI{10}{kV} and a beam current of \SI{120}{pA} was used to balance spatial resolution and X-ray signal strength. Elemental mapping was conducted using L\textsubscript{$\alpha$} X-ray lines for gallium (Ga) and M\textsubscript{$\zeta$} lines for indium (In) and tin (Sn) to ensure optimal signal separation and accuracy in quantifying the distribution of these metallic elements.

\begin{figure*}
    \centering
    \includegraphics[width=0.7\textwidth]{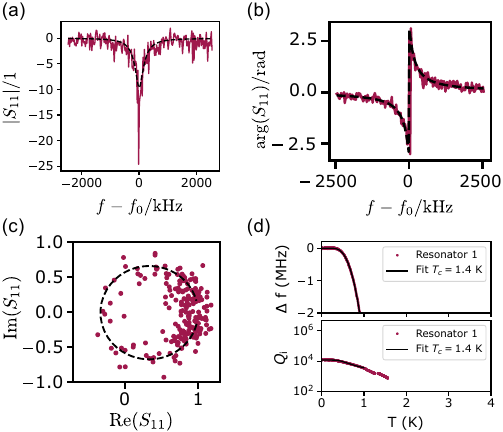} 
    \caption[Microwave measurement of Resonator 1 during second cool-down.]{\textbf{Microwave measurement of Resonator 1 during second cool-down.}
    \textbf{(a)} Amplitude, \textbf{(b)} phase of the reflection coefficient close to the resonance frequency of the resonator and single average photon number excitation. In comparison to the first cool-down, the resonance shifted toward higher frequencies ($f_0 = \SI{6.18}{GHz}$). The much larger FWHM of the resonance is mainly due to the increased coupling of the resonator in the waveguide at this frequency, which is close to the \SI{6}{GHz}-box-mode of the waveguide. \textbf{(c)} The complex reflection coefficient indicates that the resonator is close to being critically coupled, as the circle crosses the real axis close to the origin. The internal quality factor is thus also smaller and close to the coupling quality factor of $1.5 \times 10^4$. However, the stronger coupling likely caused the waveguide surfaces to limit internal quality factors. This is supported (see text) by the temperature dependent frequency shift $\Delta f(T) = f(T)-f_0$ and internal quality factor $Q_\mathrm{i}(T)$ shown in \textbf{(d)}. 
    } 
    \label{fig:R1CD2}
\end{figure*}

\section{Microwave measurements during second cool-down}
Microwave measuremtns during the second cool-down of Resonator 1 are depicted in \hyperref[fig:R1CD2]{FIG.~11}. Compared to the initial cool-down presented in the main section of this paper, the amplitude profile exhibits a significantly greater depth and broader width, suggesting a reduced internal quality factor.  The circle fit in the complex plane presents an internal quality factor that decreased by over one order of magnitude (from $5 \times 10^5$ to $1.5 \times 10^4$). 
The thermal quasi-particle model fit to the temperature dependence of frequency and quality factor, reveals that $T_\mathrm{c} = \SI{1.4}{K}$, which is close to the critical temperature of the waveguide material aluminum ($T_\mathrm{c} = \SI{1.2}{K}$).

\section{Printed resonators on different substrates}
\begin{figure*}
    \centering
    \includegraphics[width=\textwidth]{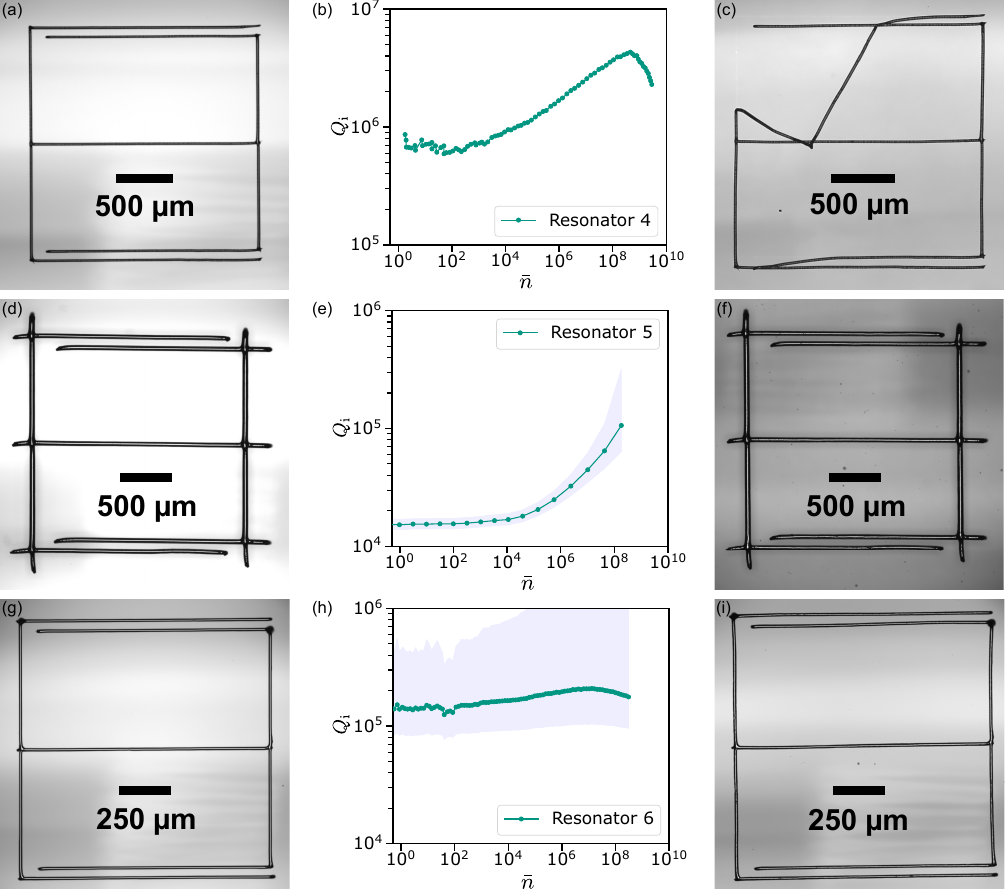}
    \caption[Printed resonators on silicon and MgO before and after $Q_\mathrm{i}$ measurements.]{\textbf{Printed resonators on silicon and MgO before and after $Q_\mathrm{i}$ measurements.}
    \textbf{(a)} Optical micrograph of an EGaInSn resonator printed on intrinsic silicon before cool-down (Resonator 4). 
    \textbf{(b)} $Q_\mathrm{i}$ measurement of Resonator 4 ($f_0 = \SI{5.593}{GHz}$) as a function of microwave power given as average photon number $\bar{n}$. The observed $Q$-factor is approaching 1 million at lower microwave powers.
    \textbf{(c)} Microscope image of the resonator after warm-up. The top left corner and a portion of the lower capacitor trace detached from the silicon substrate, nevertheless, the structural integrity was maintained.
    \textbf{(d)} Optical micrograph of an EGaInSn resonator printed on magnesium oxide (MgO) substrate before cool-down (Resonator 5).
    \textbf{(e)} $Q_\mathrm{i}$ plotted against the average photon number $\bar{n}$. The internal quality factor at $f_0 = \SI{5.724}{GHz}$ is more than one order lower than the ones printed on sapphire and silicon. This is most likely related to the higher substrate losses of MgO in comparison to sapphire and silicon.
    \textbf{(f)} Microscope image of the MgO resonator after warm-up. The structural integrity was maintained in its entirety.
    \textbf{(g)} Optical micrograph of a printed 1.3 mm × 1.3 mm EGaInSn resonator on silicon (Resonator 6) before first cool-down. 
    \textbf{(h)} $Q_\mathrm{i}$ measurement of Resonator 6 at a resonant frequency of $f_0 = \SI{8.441}{GHz}$. Internal quality factor measurements are in the range of $1.5 \times 10^5$ at lower powers. The shaded region indicates the uncertainty in the $Q_\mathrm{i}$ measurement stemming from Fano interference.
    \textbf{(i)} Optical micrograph of Resonator 6 after two full cryo cycles. The reflectivity on some areas of the structure changed. Also, the vertical lines bent slightly and the capacitor gap at the top widened marginally after multiple temperature cycles.
    } 
    \label{fig:Substrates}
\end{figure*}

In addition to the resonators printed on sapphire, other substrate materials and resonator dimensions were investigated as displayed in \hyperref[fig:Substrates]{FIG.~12}. Most notably, Resonator 4 fabricated on pure silicon exhibits promising microwave measurement results, with an internal quality factor $Q_\mathrm{i}$ comparable to that of resonators on sapphire. Unfortunately, the resonator was also found degraded after temperature cycling. Interestingly, the mode of failure appears to be different than what was observed on sapphire substrates: The structure remains largely intact, with only partial delamination from the substrate of the lower capacitor line and the top-left corner. Improved surface adhesion, for example through oxygen plasma cleaning, may mitigate this issue in the future. 

Resonator 5, fabricated on a magnesium oxide substrate, exhibits a lower internal quality factor than those on sapphire and silicon. However, its structural stability appears superior to that of the previous resonator, with minimal changes observed in microscopic images. Plasma cleaning of MgO substrates was used before printing instead of cleaning in deionized water, as MgO reacts strongly with water. 

A scaled-down design was developed to investigate the dependence of loss on interface participation. It was fabricated on a silicon substrate. The $1.3$~mm × $1.3$~mm Resonator 6 exhibits internal quality factors on the order of $1.5 \times 10^5$ at single-photon power levels, which indicates that the resonators are limited by dielectric loss occurring at the surface or interface with the substrate. The device was fabricated using the same process as Resonator 4 (without oxygen plasma treatment), however, the degradation observed after cryogenic cycling is less pronounced than in the $2$~mm × $2$~mm  sample.

\begin{figure*}
    \centering
    \includegraphics[width=0.5\textwidth]{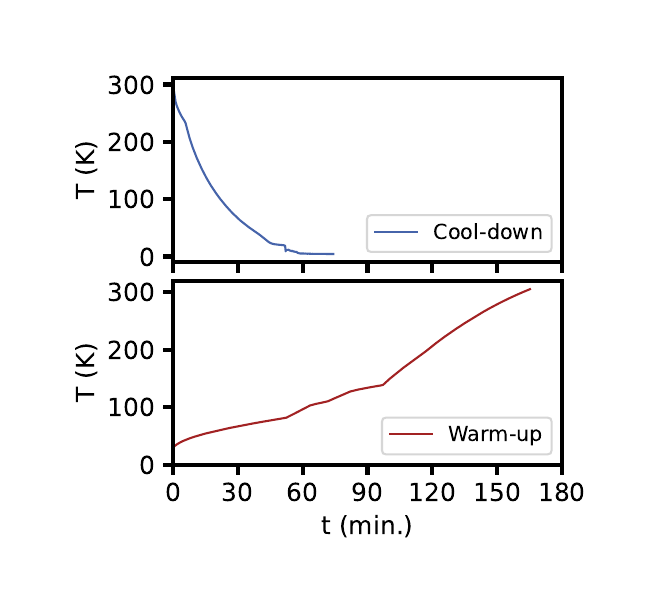}
    \caption[Temporal evolution of the cryostat temperature for the optical cryo-microscopy experiment.]{\textbf{Temporal evolution of the cryostat temperature for the optical cryo-microscopy experiment.} The temperature of the cryo-probestation sample stage was measured using a Lake Shore 336 cryogenic temperature controller, which recorded the temperature at one-second intervals. Note that the helium flow was manually controlled, which accounts for the slight fluctuations observed in the cool-down graph. The warm-up was performed using the integrated heater of the cryostat. 
    } 
    \label{fig:T_vs_t}
\end{figure*}

\section{Optical cryo-microscopy}

\hyperref[fig:T_vs_t]{FIG.~13} documents the temperature as a function of time during the optical microscopy experiment. The upper plot of the temperature during cool-down exhibits slight fluctuations, likely due to the manual control of helium flow in the cryostat. The change in slope observed in the warm-up curve after \SI{100}{min} most likely corresponds to an increase in heater power to accelerate the experiment. Note that the cool-down and warm-up times in the dilution cryostat used for the microwave measurements were much slower, which likely is the reason for why the resonators only break during warm-up there and not already during cool-down.

\clearpage
\bibliography{apssamp}

\end{document}